\documentclass[preprint,superscriptaddress,amsmath,amssymb, floatfix,aps,pre, showpacs]{revtex4}

\usepackage{graphics,graphicx}
\usepackage{bm}

\usepackage[usenames]{color}

\begin{document}

\title{Thermal Excitations of Warped Membranes}

\author{Andrej Ko\v{s}mrlj}
\email{andrej@physics.harvard.edu}
\affiliation{Department of Physics, Harvard University, Cambridge, MA 02138}

\author{David R. Nelson}
\affiliation{Department of Physics, Harvard University, Cambridge, MA 02138}
\affiliation{Department of Molecular and Cellular Biology, and School of Engineering and Applied Science, Harvard University, Cambridge, Massachusetts 02138}

\begin{abstract}

We explore thermal fluctuations of thin planar membranes with a frozen spatially-varying background metric and a shear modulus. We focus on a special class of $D$-dimensional ``warped membranes'' embedded in a $d-$dimensional space with $d\ge D+1$ and a preferred height profile characterized by quenched random Gaussian variables $\{h_\alpha({\bf q})\}$, $\alpha=D+1,\ldots, d$, in Fourier space with zero mean and a power law variance $\overline{ h_\alpha({\bf q}_1) h_\beta({\bf q}_2) } \sim \delta_{\alpha, \beta} \, \delta_{{\bf q}_1, -{\bf q}_2} \, q_1^{-d_h}$. The case $D=2$, $d=3$ with $d_h = 4$ could be realized by flash polymerizing lyotropic smectic liquid crystals.
For $D < \max\{4, d_h\}$ the elastic constants are non-trivially renormalized and become scale dependent. Via a self consistent screening approximation we find that the renormalized bending rigidity increases for small wavevectors  ${{\bf q}}$ as $\kappa_R \sim q^{-\eta_f}$, while the in-hyperplane elastic constants decrease according to $\lambda_R,\ \mu_R  \sim q^{+\eta_u}$. The quenched background metric is relevant (irelevant) for warped membranes characterized by exponent $d_h > 4 - \eta_f^{(F)}$ ($d_h < 4 - \eta_f^{(F)}$), where $\eta_f^{(F)}$ is the scaling exponent for tethered surfaces with a flat background metric, and the scaling exponents are related through $\eta_u + \eta_f = d_h - D$ ($\eta_u + 2 \eta_f=4-D$).

\end{abstract}
\maketitle

\section{Introduction}

Thermal fluctuations strongly affect the long-wavelength elastic properties of thin tethered membranes, giving rise to scale dependent elastic moduli. Due to the interplay between the local stretching and bending, the macroscopic bending rigidity diverges at long wavelengths, while the bulk and shear moduli tend to zero. These remarkable effects have been measured experimentally through the flickering of red blood cells~\cite{schmidt93} and in a number of numerical studies~\cite{zhang93, zhang96, bowick96, paulose12}. Membranes of general shapes are hard to treat analytically, but progress is possible for simple flat surfaces~\cite{nelsonB, paczuski88, aronovitz89, guitter89, ledoussal92} even in a presence of quenched disorder (e.g. quenched random metric or curvature)~\cite{radzihovsky91, morse92, radzihovsky92, ledoussal93, kosmrlj13}. Quenched disorder can result in a scale dependent elastic properties even in the absence of thermal fluctuations at zero temperature~\cite{radzihovsky91, morse92, radzihovsky92, ledoussal93, kosmrlj13}.

In this paper we study how thermal excitations renormalize the elastic properties of a particularly simple ``unfrustrated'' class of  nearly flat quenched random $D$-dimensional membranes embedded in $d$-dimensional space, which we call ``warped membranes''~\cite{kosmrlj13}. In the Monge representation a preferred membrane configuration ${\bf X}^0(x^k)$ is described with a random height profile $h_\alpha (x^k)$ such that
\begin{equation}
{\bf X}^0= x^i \hat {\bf e}_i + h_\alpha \hat {\bf e}_\alpha,
\label{eq:ref_state}
\end{equation}
where we use the convention of summing over repeated indices unless otherwise stated.
Here, $\{\hat e_i, \hat e_\alpha\}$ are orthonormal Euclidian vectors and Latin and Greek indices run from $1,\ldots,D$ and from $D+1,\ldots, d$ respectively. As discussed in Ref.~\cite{kosmrlj13}, this system lacks geometrical frustration, similar to Mattis models of random spin systems~\cite{mattis76}, and hence is particularly simple to analyze. In Fourier space the quenched height profiles $h_\alpha({\bf q}) =\int \! d^D{\bf x}\, e^{-i {\bf q} \cdot {\bf x}} h_\alpha({\bf x})/A$ are assumed to be independent random Gaussian variables with zero mean and a power law variance for small ${\bf q}$,
\begin{equation}
\overline{h_\alpha({\bf q}_1) h_{\beta}({\bf q}_2)} = \delta_{\alpha,\beta} \delta_{{\bf q}_1, - {\bf q}_2} \frac{\Delta^2}{A q_1^{d_h}} \equiv  \delta_{\alpha,\beta} \delta_{{\bf q}_1, - {\bf q}_2} G_{hh}({\bf q}_1),
\label{eq:h_variance}
\end{equation}
where $\delta$ is the Kronecker's delta, $A$ is a $D$-dimensional projected area of the membrane, and the overbar denotes averaging over a quenched random Gaussian probability distribution.

In a previous publication~\cite{kosmrlj13} we described how $D=2$-dimensional warped membranes characterized by undulation exponents $d_h=4$, $2$ and $0$ could be realized experimentally by flash polymerizing thermally fluctuating flat lipid bilayers, or by using a rough surface of the crystal as a membrane template. There we focused on mechanical properties of such warped membranes embedded in $d=3$-dimensional space  at $T=0$ (i.e., in the absence of thermal fluctuations) and demonstrated, by using the self-consistent screening approximation,~\cite{bray74a, bray74b} that elastic constants become scale dependent for membranes with $d_h > 2$ and scale as $\kappa_R \sim q^{-\eta_f}$ and $\lambda_R, \mu_R \sim q^{+\eta_u}$, with $\eta_f = \eta_u = (d_h - 2)/2$. In this paper we generalize our results to abstract membranes in an arbitrary number of dimensions and include the effects of thermal fluctuations to predict the scaling exponents $\eta_u$ and $\eta_f$. We show that for membranes characterized by $d_h < 4 - \eta_f^{(F)}$ the quenched random background metric is irrelevant and the exponents $\eta_u$ and $\eta_f$ have the same value as those for the thermally fluctuating surfaces with a \emph{flat} (F) background metric. The quenched background metric becomes relevant for $d_h \ge 4 - \eta_f^{(F)}$ and changes the exponents of the scale-dependent elastic constants.

The rest of the paper is organized as follows. In Sec.~\ref{sec:free_energy} we discuss the free energy cost of warped membrane deformations, which are decomposed into in-hyperplane deformations and out-of-hyperpane deformations. In Sec.~\ref{sec:correlation_functions} we introduce the correlation functions of such deformations and make connections to the scale dependent elastic constants. Finally, in Sec.~\ref{sec:SCSA} we use the self-consistent screening approximation to estimate the scaling exponents $\eta_f$ and $\eta_u$ that describe these quantities at long wavelengths.

\section{Free energy cost of thin membrane deformations}
\label{sec:free_energy}
Deformations of a nearly flat reference warped membrane are described by mapping a configuration ${\bf X}^0(x^k)$ into a configuration ${\bf X}(x^k)$, together with an associated free energy cost.
For small deformations the free energy cost of deformation~\cite{nelsonB, radzihovsky91, morse92, ledoussal93, kosmrlj13} can be expressed as 
\begin{equation}
F[{\bf X}] = \int \!\! d^D{\bf x}\, \frac{1}{2} \left[\lambda u_{ii}^2 + 2 \mu u_{ij}^2 + \kappa {\bf K_{ii}}^2 \right],
\label{eq:free_energy}
\end{equation}
where $u_{ij}(x^k)$ and ${\bf K}_{ij}(x^k)$ are respectively the local strain and the local bending strain tensors and are defined as
\begin{equation}
u_{ij}=(\partial_i {\bf X} \cdot {\partial_j} {\bf X} - A_{ij})/2, \qquad {\bf K}_{ij} = \partial_i \partial_j {\bf X} - {\bf B}_{ij}.
\end{equation}
Here, $A_{ij}(x^k)$ and ${\bf B}_{ij}(x^k)$ are quenched random matrices that arise from the preferred local metric and the curvature tensors respectively. For arbitrary $A_{ij}$ and ${\bf B}_{ij}$ there is in general no membrane configuration ${\bf X}(x^k)$ that would correspond to the zero free energy in Eq.~ (\ref{eq:free_energy}). However, a unique ground state without strains is in fact possible when these quenched tensors satisfy the Gauss-Codazzi-Mainardi relations,~\cite{spivakB} and can thus be expressed in terms of the metric tensor $A_{ij}=\partial_i {\bf X}^0 \cdot {\partial_j} {\bf X}^0$ and the curvature tensor ${\bf B}_{ij}=\partial_i \partial_j {\bf X}^0$ of a preferred membrane configuration ${\bf X}^0(x^k)$ that corresponds to the minimum free energy. Such an unfrustrated model resembles Mattis models of spin glasses~\cite{mattis76}. The mechanical properties of these ``warped membranes'' at $T=0$ were discussed in Ref.~\cite{kosmrlj13}.

Thermal fluctuations of membranes in a presence of \emph{independent} quenched random tensors $A_{ij}(x^k)$ and ${\bf B}_{ij}(x^k)$ have been studied before and it was shown that quenched averaged renormalized elastic constants can become length scale dependent,~\cite{radzihovsky91, morse92, radzihovsky92, ledoussal93} with scaling exponents that differ from those for flat surfaces ($A_{ij}=\delta_{ij}$, ${\bf B}_{ij}={\bf 0}$). In this paper we study the effect of thermal fluctuations on a particular class of quenched random tensors, which are no longer independent and correspond exactly to the metric tensor $A_{ij}$ and the curvature tensor ${\bf B}_{ij}$ of the preferred random membrane configuration ${\bf X}^0 (x^k)$ displayed in Eq.~(\ref{eq:ref_state}), i.e.
\begin{equation}
A_{ij}=\partial_i {\bf X}^0 \cdot {\partial_j} {\bf X}^0, \qquad {\bf B}_{ij} = \partial_i \partial_j {\bf X}^0.
\end{equation}

Deformations of membranes are typically decomposed into in-hyperplane displacements $u_i(x^k)$ and out-of-hyperplane displacements $f_\alpha(x^k)$, such that
\begin{equation}
{\bf X} = {\bf X}^0 + u_i \hat{\bf t}_i + f_\alpha \hat{\bf n}_\alpha,
\end{equation}
where $\hat {\bf t}_i = (\hat {\bf e}_i + \sum_\alpha (\partial_i h_\alpha) \hat{\bf e}_\alpha)/\sqrt{1 + \sum_\alpha (\partial_i h_\alpha)^2}$ are local tangent vectors and $\hat {\bf n}_\alpha = (\hat{\bf e}_\alpha - \sum_i (\partial_i h_\alpha) \hat{\bf e}_i)/\sqrt{1 + \sum_i (\partial_i h_\alpha)^2}$ are local normal vectors. In this decomposition the local strain tensor $u_{ij}(x^k)$ and bending strain tensor ${\bf K}_{ij}(x^k)$ become
\begin{eqnarray}
u_{ij}&=& \frac{1}{2} \left( \partial_i u_j + \partial_j u_i \right) + \frac{1}{2} (\partial_i f_\alpha) (\partial_j f_\alpha) - f_\alpha \partial_i \partial_j h_\alpha, \nonumber \\
{\bf K}_{ij} &=& (\partial_i \partial_j f_\alpha) \hat{\bf e}_\alpha, 
\label{eq:strain_tensor}
\end{eqnarray}
where we kept only the lowest order terms in $u_i$, $f_\alpha$ and $h_\alpha$. As discussed in detail in Ref.~\cite{kosmrlj13} for the case $T=0$, $d=3$, $D=2$, this parameterization generalizes shallow shell theory~\cite{koiterB, novozhilovB,sanders63} for arbitrary nearly flat membranes with $D$ internal dimensions embedded in an external space of dimension $d$.

\section{Correlation functions}
\label{sec:correlation_functions}
By adapting the fluctuation-response theorems of statistical mechanics,~\cite{callen51} the effective elastic properties of warped membranes can be extracted from appropriate correlation functions. To explain our procedure, we first introduce the harmonic approximation, where we keep only the first term in the strain tensor $u_{ij}$ in Eq.~(\ref{eq:strain_tensor}). In this harmonic approximation the in-hyperplane deformations $u_i$ and out-of-hyperplane deformations $f_\alpha$ are decoupled and in the Fourier space the free energy can be expressed as
\begin{eqnarray}
\frac{F_0}{A}\! &=& \!\frac{\lambda}{2} {(u_{ii}^0)}^2 + \mu {(u_{ij}^0)}^2 + \sum_{\bf q} \frac{\kappa q^4}{2}  f_\alpha ({\bf q}) f_{\alpha} (-{\bf q})  \nonumber \\
&& + \sum_{\bf q} \frac{q^2}{2} u_i ({\bf q})\!\left[(2 \mu + \lambda) P^L_{ij}({\bf q}) + \mu P^T_{ij} ({\bf q}) \right] \!u_j (-{\bf q}),
\end{eqnarray}
where $A$ is a $D$-dimensional projected area of the membrane. Here, we separated out the uniform strain $u_{ij}^0$, and introduced the longitudinal and transverse projector operators $P^L_{ij}({\bf q}) = q_i q_j/q^2$ and $P^T_{ij} ({\bf q}) = \delta_{ij} - q_i q_j/q^2$, which decouple the in-hyperplane displacements $u_i({\bf q})$ into one longitudinal mode $u_L ({\bf q})$ and $D-1$ orthogonal transverse modes $u^a_T({\bf q})$, $a=1,\ldots,D-1$. In the harmonic approximation the correlation functions are\begin{eqnarray}
\left<f_{\alpha}({\bf q}_1) f_\beta ({\bf q}_2) \right>_0 \!&=&\! \delta_{\alpha,\beta} \delta_{{\bf q}_1, -{\bf q}_2} \frac{k_B T}{A \kappa q^4},\nonumber \\
\left<u_L({\bf q}_1) u_L ({\bf q}_2) \right>_0 \!&=&\! \delta_{{\bf q}_1, -{\bf q}_2} \frac{k_B T}{A (2 \mu + \lambda) q^2}, \nonumber \\
\left<u^a_T({\bf q}_1) u^b_T ({\bf q}_2) \right>_0 \!&=&\! \delta_{a,b} \delta_{{\bf q}_1, -{\bf q}_2} \frac{k_B T}{A \mu q^2},
\label{eq:harmonic_correlation_functions}
\end{eqnarray}
where $a$ and $b$ correspond to indices of $D-1$ transverse in-hyperplane modes and brackets denote thermal averaging $\left<\mathcal O \right>_0 =\int\! \mathcal D[u_i, f_\alpha] \,\mathcal O e^{-F_0/k_B T}/Z_0$, where
$Z_0 ={\int \!\mathcal D[u_i, f_\alpha] \,e^{-F_0/k_B T}}$ is the partition function in the harmonic approximation.

As we shall see, the nonlinear couplings between the in-hyperplane displacements $u_i$, the out-of-hyperplane displacements $f_\alpha$, and the quenched random background metric $h_\alpha$, through the strain tensor $u_{ij}$ in Eq.~(\ref{eq:strain_tensor}), effectively renormalize the correlation functions at long wavelengths (${\bf q} \rightarrow {\bf 0}$) to
\begin{eqnarray}
\overline{\left<f_\alpha({\bf q}_1) f_\beta({\bf q}_2)\right>_c} & \equiv & \delta_{\alpha,\beta} \delta_{{\bf q}_1, -{\bf q}_2} \ G_{ff} ({\bf q}_1) \ \,\sim q_1^{-4+\eta_f}, \nonumber \\
\overline{\left<u_L({\bf q}_1) u_L({\bf q}_2)\right>_c} & \equiv & \phantom{\delta_{a,b} }\delta_{{\bf q}_1, -{\bf q}_2} G_{u_L u_L} ({\bf q}_1) \sim q_1^{-2-\eta_{u_L}}, \nonumber \\
\overline{\left<u^a_L({\bf q}_1) u^b_L({\bf q}_2)\right>_c} & \equiv & \delta_{a,b} \delta_{{\bf q}_1, -{\bf q}_2} G_{u_T u_T} ({\bf q}_1) \sim q_1^{-2-\eta_{u_T}}.
\label{eq:correlation_functions}
\end{eqnarray}
In the expressions above we first perform the thermal averaging denoted with brackets $<>$ (as above but with the free energy $F$ replacing the $F_0$)
and the subsctipt $c$ corresponds to the cummulants or ``connected'' averages $\left< \mathcal{A} \mathcal{B} \right>_c = \left< \mathcal{A} \mathcal{B} \right> - \left< \mathcal{A} \right> \left< \mathcal{B} \right>$.
The second averaging denoted with the overbar $\overline {\mathcal O} = \int\! \mathcal D[h_\alpha] \, \mathcal O \, \mathcal  P(h_\alpha)$
 is done over the quenched random background metric, with $\mathcal P(h_\alpha)$ being a Gaussian distribution with $0$ mean and the variance described in Eq.~(\ref{eq:h_variance}). In comparison with correlation functions within the harmonic approximation in Eq.~(\ref{eq:harmonic_correlation_functions}), which we refer to as $G^0_{ff}$, $G^0_{u_L u_L}$, and $G^0_{u_T u_T}$ in the rest of the paper, the scaling behaviour of the correlation functions in Eq.~(\ref{eq:correlation_functions}) can be interpreted as a scale dependent (i.e., wave vector dependent) effective elastic constants $\kappa_R({\bf q}) \sim q^{-\eta_f}$, $2 \mu_R({\bf q}) + \lambda_R({\bf q}) \sim q^{+\eta_{u_L}}$ , $\mu_R({\bf q})\sim q^{+\eta_{u_T}}$. For the rest of the paper our main goal is to determine the scaling exponents $\eta_f$, $\eta_{u_L}$ and $\eta_{u_T}$.
 
In addition to the cumulants we will also discuss the ``disconnected'' averages
\begin{eqnarray}
\overline{\left<f_\alpha({\bf q}_1)\right>\left< f_\beta({\bf q}_2)\right>} & \equiv & \delta_{\alpha,\beta} \delta_{{\bf q}_1, -{\bf q}_2} \ G'_{ff} ({\bf q}_1) \ \,\sim q_1^{-4+\eta'_f}, \nonumber \\
\overline{\left<u_L({\bf q}_1)\right>\left< u_L({\bf q}_2)\right>} & \equiv & \phantom{\delta_{a,b} }\delta_{{\bf q}_1, -{\bf q}_2} G'_{u_L u_L} ({\bf q}_1) \sim q_1^{-2-\eta'_{u_L}}, \nonumber \\
\overline{\left<u^a_L({\bf q}_1)\right>\left< u^b_L({\bf q}_2)\right>} & \equiv & \delta_{a,b} \delta_{{\bf q}_1, -{\bf q}_2} G'_{u_T u_T} ({\bf q}_1) \sim q_1^{-2-\eta'_{u_T}}.
\label{eq:disconnected_correlation_functions}
\end{eqnarray}
Note that for tethered surfaces that are flat in the ground state, such averages vanish.

Since the correlaction functions cannot be evaluated exactly, we have to make certain approximations. Perturbation expansions in temperature $T$~\cite{nelsonB, radzihovsky91} and in quenched disordered metric amplitude $\Delta$ (see Eq.~(\ref{eq:h_variance}))~\cite{kosmrlj13} converge only when the membrane dimensionality is $D>\max\{4,d_h\}$. In this case the elastic constants have a finite renormalization and are scale independent ($\eta_f=\eta_{u_T}=\eta_{u_L}=0$) for long wavelengths (small $q$), as in conventional elasticity theory. The more interesting case is when $D<\max\{4,d_h\}$ and the elastic constants become scale-dependent at long wavelengths; this case is considered in the rest of the paper. Because the perturbation expansion diverges, we use the self-consistent screening approximation (SCSA)~\cite{bray74a, bray74b} to approximate the scaling exponents $\eta_f$, $\eta_{u_L}$, and $\eta_{u_T}$. Note that the quenched averaged properties can sometimes be calculated using the replica trick~\cite{mezardB}. However, the approach taken here avoids the usual zero-replica limit, by recognizing that the SCSA method evaluates an infinite subset of all terms in the perturbation expansion (e.g., see~\cite{kosmrlj13}).

\section{Self-consistent screening approximation}
\label{sec:SCSA}

The SCSA was first introduced to estimate critical exponents in the Landau-Ginzburg model of critical phenomena~\cite{bray74a, bray74b} and was later applied to calculate the effective elastic constants due to thermal fluctuations of flat tethered surfaces~\cite{ledoussal92, gazit09, zakharchenko10} and also to study their properties in the presence of quenched random disorder~\cite{radzihovsky92,ledoussal93,kosmrlj13}. For thermally fluctuating flat tethered membranes, the SCSA method~\cite{ledoussal92, gazit09, zakharchenko10} gives more accurate scaling of elastic constants than the first order epsilon expansion in renormalization group~\cite{aronovitz88,aronovitz89}. The SCSA method is equivalent to a $1/(d-D)$ expansion and thus becomes exact when the embedding space dimension $d$ is large compared to the manifold dimension $D$.

First we focus on the out-of-hyperplane displacements $f_\alpha$. The algebra is greatly simplified if we initially integrate out the in-hyperplane displacements $u_i$ by~\cite{nelsonB}
\begin{equation}
e^{-F_\textrm{eff}/k_B T} \equiv \int\! \mathcal D [u_i] e^{-F/k_B T},
\end{equation}
 which leads to the effective free energy
 \begin{eqnarray}
 \frac{F_\textrm{eff}}{A} &=& \sum_{\bf q} \frac{\kappa q^4}{2} f_{\alpha} ({\bf q}) f_{\alpha} (-{\bf q}) + \sum_{\substack{{\bf q}_1 + {\bf q}_2 = {\bf q} \ne {\bf 0}\\{\bf q}_3 + {\bf q}_4 = -{\bf q}\ne{\bf 0}}} S_{ij} ({\bf q}_1, {\bf q}_2) R^0_{ij,kl}({\bf q}) S_{kl} ({\bf q}_3, {\bf q}_4).
 \end{eqnarray}
Here, we introduced the vertex $R^0_{ij,kl}$ and the tensor $S_{ij}$,
\begin{eqnarray}
R^0_{ij,kl} \!&=&\! \frac{\mu}{2} \left(P^T_{ik}P^T_{jl} +P^T_{il} P^T_{jk} + \frac{2 \lambda}{(2 \mu+\lambda)} P^T_{ij} P^T_{kl} \right),\nonumber \\
S_{ij}({\bf q}_1, {\bf q}_2) \!&=&\! \frac{1}{2} q_{1i}  q_{2j} f_{\alpha}({\bf q}_1) f_\alpha({\bf q}_2) - q_{2i} q_{2j} f_{\alpha} ({\bf q}_1) h_\alpha({\bf q}_2), 
\end{eqnarray}
where we again use the transverse projector operator $P^T_{ij}({\bf q}) = \delta_{ij}-q_i q_j/q^2$. Note that the vertex $R^0_{ij,kl}$ can be rewritten~\cite{ledoussal92} as $R^0_{ij,kl}= \mu M_{ij,kl} + \rho N_{ij,kl}$, where
\begin{eqnarray}
N_{ij,kl} &=& \frac{1}{(D-1)} P_{ij}^T P_{kl}^T, \nonumber \\
M_{ij,kl} & = & \frac{1}{2} \left( P^T_{ik} P^T_{jl} + P^T_{il} P^T_{jk} \right) - N_{ij,kl},
\end{eqnarray}
where $\mu$ is the shear modulus and $\rho=\mu (2 \mu + D \lambda)/(2 \mu + \lambda)$. The convenience of this decomposition is that $M$ and $N$ are mutually orthogonal under matrix multiplicaion (e.g. $M_{ij,kl}M_{kl,mn}=M_{ij,mn}$, $M_{ij,kl}N_{kl,mn}=0$, etc.).

\begin{figure}
\includegraphics[scale=1]{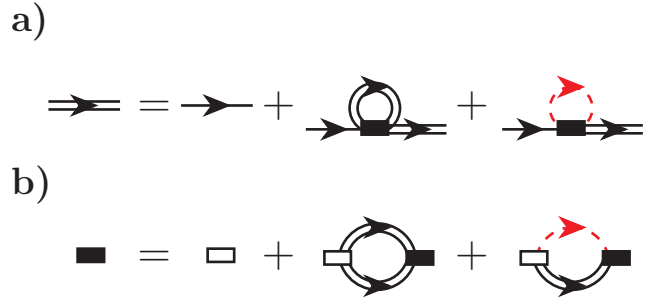}
\caption{(Color online) Schematic description of the SCSA method for a) the renormalized propagator $G_{ff}$ and b) the renormalized vertex $R$. Single and double solid lines represent the bare propagator $G^0_{ff}$ and the renormalized propagator $G_{ff}$ respectively, red dashed lines represent the quenched disorder propagator $G_{hh}$, and empty and solid rectangles correspond to the bare vertex $R^0$ and the renormalized vertex $R$ respectively.}
\label{fig:scsa_scheme}
\end{figure}

The SCSA approximation is schematically presented in Fig.~\ref{fig:scsa_scheme}, which summarizes a set of coupled integral equations for the renormalized propagator $G_{ff}({\bf q})$ and for the renormalized vertex $R_{ij,kl}({\bf q})$, namely
\begin{eqnarray}
\frac{1}{G_{ff}({\bf q})} & = & \frac{1}{G^0_{ff}({\bf q})} + \frac{2 A}{k_B T} \sum_{\bf p} \left[G_{ff}({\bf q} + {\bf p}) + G_{hh} ({\bf q} + {\bf p}) \right] q_i q_j  R_{ij,kl}({\bf p}) q_k q_l, \nonumber\\
R_{ij,kl} ({\bf q}) & = & R^0_{ij,kl} ({\bf q}) - \frac{(d-D) A}{k_B T} \sum_{\bf p} G_{ff} ({\bf q}+ {\bf p}) \left[G_{ff}({\bf p}) + 2 G_{hh} ({\bf p})\right] R^0_{ij,mn} ({\bf q}) p_m p_n p_r p_s R_{rs,kl}({\bf q}).\nonumber \\
\label{eq:scsa_equations}
\end{eqnarray}
The matrix equation for the renormalized vertex $R$ above can be solved in terms of the renormalized elastic constants for orthogonal matrices $M$ and $N$
\begin{eqnarray}
\frac{1}{\mu_R({\bf q})} &=& \frac{1}{\mu} + 2 \Pi({\bf q}), \nonumber \\
\frac{1}{\rho_R({\bf q})} &=& \frac{1}{\rho} + (D+1) \Pi({\bf q}),
\label{eq:mu_rho}
\end{eqnarray}
where $\Pi({\bf q})$ is
\begin{eqnarray}
\Pi({\bf q})\!&=&\!\frac{(d-D) A}{k_B T (D^2-1)} \sum_{\bf p} (p_i P^T_{ij}({\bf q}) p_j)^2 G_{ff} ({\bf q} + {\bf p}) \left[G_{ff}({\bf p}) + 2 G_{hh} ({\bf p})\right].
\label{eq:pi}
\end{eqnarray}
The set of integral equations above can be solved self-consitently by assuming that $G_{ff}({\bf q})=C_f q^{-4+\eta_f}/A$ and deriving a self-consistent equation for $\eta_f$. Insertion of this power law ansatz into Eq.~(\ref{eq:pi}) leads to
\begin{eqnarray}
\Pi ({\bf q})&=& \frac{(d-D) C_f}{k_B T (D^2 -1)} \left[\frac{C_f I\left(2-\frac{\eta_f}{2},2-\frac{\eta_f}{2} \right)}{q^{4-D-2 \eta_f}}  + \frac{2 \Delta^2 I\left(2-\frac{\eta_f}{2},\frac{d_h}{2} \right)}{q^{d_h-D-\eta_f}} \right],
\label{eq:pi2}
\end{eqnarray}
where we introduced the function $I$ through the integral
\begin{equation}
\frac{I(\alpha, \beta)}{q^{2 (\alpha+ \beta)-4-D}} =  \sum_{\bf p} \frac{(p_i P^T_{ij}({\bf q}) p_j)^2}{A p^{2 \alpha} |{\bf q} + {\bf p}|^{2 \beta}}.\end{equation}
The integral above can be evaluated exacly and can be expressed in terms of Gamma functions
\begin{eqnarray}
I(\alpha, \beta) & = & \frac{\Gamma(2+D) \Gamma(\alpha + \beta - 2-D/2)\Gamma(2+D/2 - \alpha) \Gamma(2+D/2-\beta)}{2^{(2 D + 1)} \Pi ^{(D-1)/2} \Gamma[(D-1)/2] \Gamma[1 + D/2] \Gamma[\alpha] \Gamma[\beta] \Gamma[4+D-\alpha-\beta]}.
\end{eqnarray}
Note that the expression above is only correct for $2(\alpha+\beta)>4+D$, otherwise the integral has an ultraviolet divergence. These divergences can be treated by introducing a microscopic cutoff of order the membrane thickness; the resulting integral has a finite $q$-independent value as ${\bf q} \rightarrow {\bf 0}$.

From Eqs.~(\ref{eq:mu_rho}) and (\ref{eq:pi2}) we notice that in the thermodynamic limit of large wavelengths the renormalized elastic constants scale as $\mu_R({\bf q}), \rho_R({\bf q}) \sim q^{+\eta_u}$ (the subscripted notation $\eta_u$ will be justified later) and there are two possible scenarios:
For $d_h <  4 - \eta_f$ the contribution from the quenched random background metric (the bubble diagram in Fig.~\ref{fig:scsa_scheme}b with a red dashed line) is negligible and the renormalized elastic constants scale as $\eta_u = 4-D-2\eta_f$, while for $d_h > 4 - \eta_f$ that contribution dominates and we find $\eta_u=d_h -D-\eta_f$.  In the first equation in Eq.~(\ref{eq:scsa_equations}) for  the renormalized propagator $G_{ff}$, the dominant contributions come from small $\bf p$ values, where we can use the asymptotic expressions for the renormalized vertex $R({\bf q})$ or equivalenty $\mu_R({\bf q})$ and  $\rho_R({\bf q})$. We find again that the quenched random background metric  (the loop diagram in Fig.~\ref{fig:scsa_scheme}a with a dashed red line) is irelevant (relevant) for $d_h<4-\eta_f$ ($d_h > 4 - \eta_f$). In both cases we find that in the long wavelength limit the dominant term on the right hand side  of the Eq.~(\ref{eq:scsa_equations}) scales as $A B(\eta_f) q^{4-\eta_f}/C_f$ and the scaling exponent $\eta_f$ is determined self-consistently by satisfying $B(\eta_f)=1$.

For the case when the quenched random background metric is irelevant ($d_h < 4-\eta_f$) we recover the results of Le Doussal and Radzihovsky in~\cite{ledoussal92} for tethered surfaces that are flat in the ground state, with $\eta_f$ determined by the equation
\begin{equation}
1 = \frac{D(D-1)}{d_c} \frac{I\left(2-\frac{\eta_f}{2} , \eta_f + \frac{D}{2} \right)}{I\left(2-\frac{\eta_f}{2},2-\frac{\eta_f}{2}\right)},
\end{equation}
where $d_c = d -D$ and $\eta_u=4-D-2\eta_f$. For $D=2$-dimensional membranes, the solution of the above equation is
\begin{equation}
\eta_f = \frac{2 (\sqrt{16 - 2 d_c + d_c^2} - d_c)}{(8-d_c)},
\end{equation}
which for the physical membranes ($d_c =1$) evaluates to $\eta_f \approx 0.821$ and $\eta_u\approx 0.358$. The quenched random background metric is irelevant for $d_h < 3.18$. We mentioned before that the SCSA is equivalent to a $1/d_c$ expansion (with $d_c = d -D$), where we find~\cite{ledoussal92}
\begin{equation}
\eta_f = \frac{8 (D-1)}{d_c (D+2)} \frac{\Gamma[D]}{\Gamma[2-D/2] \Gamma[D/2]^3} + \mathcal O \left(\frac{1}{d_c^2} \right)
\end{equation}
and for $D=2$ two dimensional membranes we get $\eta_f = 2/d_c$.

When the quenched random background metric is relevant, i.e. $d_h > 4 - \eta_f$, we obtain a self-consistent equation for $\eta_f$
\begin{eqnarray}
1 = \frac{D(D-1)}{2 (d-D)} \frac{I\left(\frac{d_h}{2}, 2  - \frac{(d_h - D - \eta_f)}{2} \right)}{I \left(\frac{d_h}{2}, 2 - \frac{\eta_f}{2} \right)}
\end{eqnarray}
and $\eta_u = d_h - D - \eta_f$. For the physical membranes (with $d=3$, $D=2$) we find $\eta_f = \eta_u = (d_h - 2)/2$, which is consistent with the scaling exponents calculated earlier for the mechanical properties of warped membranes at zero temperature~\cite{kosmrlj13}. For the large embedding space dimension we find
\begin{equation}
\eta_f = \frac{D (D-1)}{d_c} \frac{\Gamma[d_h/2] \Gamma[2 + D - d_h/2]}{\Gamma[D/2] \Gamma[2 + D/2] \Gamma[(d_h - D)/2] \Gamma[2 + (D-d_h)/2]} + \mathcal O \left(\frac{1}{d_c^2} \right).
\end{equation}

\begin{figure}
\includegraphics[scale=.45]{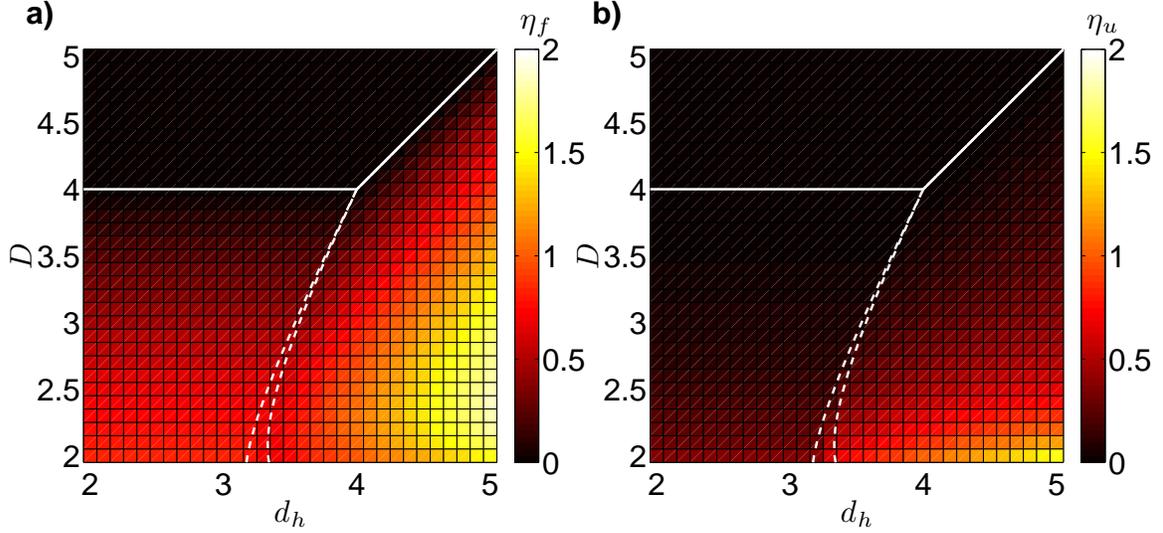}
\caption{(Color online) Heat maps for scaling exponents (a) $\eta_f$ and (b) $\eta_u$ as a function of $d_h$, the exponent characterizing a quenched random  background metric, and the membrane dimensionality $D$ embedded in $d=D+1$ dimensional space. Regions to the left (right) of the pairs of dashed white lines represent regimes where the quenched random background metric is irelevant (relevant). The space between the pairs of dashed lines correspond to the transition regions, where $\eta_f = 4 - d_h$ and $\eta_u=2d_h - 4 - D$. Note that in the top left region separated with solid white lines, i.e. for $D>\max\{4,d_h\}$, the scaling exponents are $\eta_f=\eta_u=0$.}
\label{fig:scaling_exponents}
\end{figure}

The scaling exponents $\eta_f$ and $\eta_u$ for membranes characterized with different values of $d_h$ and $D$ with $d_c=1$ are displayed in Fig.~\ref{fig:scaling_exponents}. Note the small transition region between the dashed white lines, where $\eta_f  = 4 -d_h$ and $\eta_u = 2 d_h - 4 - D$. For a related situation arising for ferromagnets with long range interactions in $4-\epsilon$ dimensions, see Ref.~\cite{sak73}.

\begin{figure}
\includegraphics[scale=1]{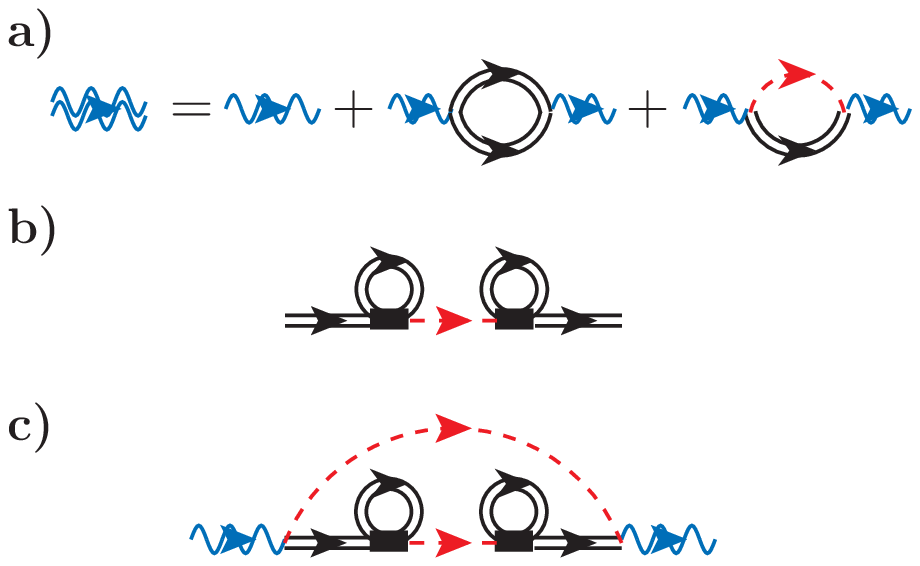}
\caption{(Color online) Schematic description for the calculations of a) the in-hyperplane correlation function $G_{uu}$ and ``disconnected'' correlation functions b) $G'_{ff}$ and c) $G'_{uu}$. The nomenclature ``disconnected'' means these graphs would be disconnected without the average over the quenched random background metric. Double solid lines represent the renormalized propagator $G_{ff}$, blue single and double wavy solid lines represent respectively the bare propagator $G^0_{uu}$  and the renormalized propagator $G_{uu}$. The red dashed lines represent the quenched disorder propagator $G_{hh}$, and solid rectangles correspond to the renormalized vertex $R$.
}
\label{fig:scsa2}
\end{figure}

We now return to the correlation functions for the \emph{in-hyperplane} displacements $u_i$. The diagrammatic representation for these correlation functions is displayed in Fig.~\ref{fig:scsa2}a; note the simillarity with the renormalized vertex function $R$ in Fig.~\ref{fig:scsa_scheme}. For both the longitudinal and transverse modes the dominant terms scale like
\begin{equation}
G_{uu} ({\bf q}) \sim \left[G^0_{uu}({\bf q}) \right]^2 \sum_{\bf p} q^2 p^4 G_{ff} ({\bf q} + {\bf p}) \left[G_{ff} ({\bf p}) + G_{hh} ({\bf p}) \right] \sim q^{-2-\eta_u},
\end{equation}
where $\eta_u$ is identical to the exponent that appeared in the renormalized vertex function. It follows that $\eta_{u_L}=\eta_{u_T}=\eta_u$. From Eq.~(\ref{eq:mu_rho}) we see that the in-hyperplane elastic constants scale like $\mu_R \sim C q^{\eta_u}/2$ and $\rho_R \sim C q^{\eta_u}/(D+1)$ or equivalently $\lambda_R \sim -C q^{\eta_u}/(D+2)$. These results suggest a universal Poisson's ratio in the long wavelength limit
\begin{equation}
\lim_{{\bf q} \rightarrow {\bf 0}} \nu_R({\bf q}) = \lim_{{\bf q} \rightarrow {\bf 0}} \frac{\lambda_R({\bf q})}{2 \mu_R ({\bf q}) + (D-1) \lambda_R({\bf q})} = -\frac{1}{3}.
\end{equation}
The numerical simulations for thermally fluctuating flat tethered surfaces~\cite{zhang96,falcioni97} (without quenched disorder) are indeed consistent with this value. However, our numerical simulations~\cite{kosmrlj13} for the mechanical properties of $D=2$-dimensional warped membranes at zero temperature suggested a \emph{positive} Poisson ratio, even though the scaling exponents were in good agreement with the SCSA predictions. This suggests that the SCSA captures the scaling exponents of elastic properties, but might be less accurate for predicting the amplitudes.

Finally, we comment on the ``disconnected'' correlation functions. The most divergent terms are sketched in Fig.~\ref{fig:scsa2} and a simple power counting gives us the scaling exponents (see Eq.~(\ref{eq:disconnected_correlation_functions}))
\begin{eqnarray}
\eta'_f &=& 2D + 2\eta_u + 4 \eta_f -4-d_h \ge \eta_f, \nonumber \\
\eta'_u & = & d_h - D - \eta'_f \le \eta_u,
\end{eqnarray}
where the equality is reached only when $\eta_f = 4 - d_h$. These results imply that the correlation functions discused before always dominate, when calculating correlation functions such as $\overline{\langle f_\alpha({\bf q}) f_\alpha(-{\bf q}) \rangle} = \overline{f_\alpha({\bf q}) f_\alpha(-{\bf q}) \rangle_c} + \overline{\langle f_\alpha({\bf q}) \rangle \langle f_\alpha(-{\bf q}) \rangle}$.

\section{Conclusions}
We have used the self-consistent screening aproximation to calculate the scalings of elastic properties for thermally fluctuating warped membranes, an especially simple class of quenched random tethered surfaces with a preordained unfrustrated ground state at $T=0$. The quenched random background metric becomes relevant and changes the scale dependence whenever $d_h \ge 4 - \eta_f^{(F)}$, where $\eta_f^{(F)}$ corresponds to the scaling exponent for the bending rigidity in the tethered surfaces with a flat background metric, and $d_h$ characterizes the scale-dependence of the quenched random disorder. That crossover can be understood heuristically as follows: In our previous study~\cite{kosmrlj13} of the zero temperature mechanical properties of warped membranes we found that the divergence of the height profile variance $\overline {|h_\alpha({\bf x})^2|} \sim L^{d_h - D}$ with the membrane size $L$ controls the scaling of elastic properties. For the thermally fluctuating tethered surfaces considered here the out-of-hyperplane displacement variance diverges as $\overline{\langle |f_\alpha({\bf x})|^2 \rangle} \sim L^{4-\eta_f-D}$. These results suggest that the quenched random background metric characterizing warped membranes becomes relevant only when the typical height fluctuations due to the frozen metric are larger than the thermal undulations for the normal out-of-hyperplane displacements.

\acknowledgements{We acknowledge support by the National Science Foundation, through grants DMR1005289 and DMR1306367 and through the Harvard Materials Research and Engineering Center through Grant DMR-0820484.}

\bibliography{library}
\end{document}